\documentclass[journal]{IEEEtran} % Removed "lettersize" option as IEEEtran defaults to letter size.

\usepackage[linesnumbered,ruled,vlined]{algorithm2e} % Only keeping algorithm2e for consistency.
\usepackage{cite}
\usepackage{amsmath, amsfonts} % Merged amsfonts into amsmath line for cleaner code.
\usepackage{array, color}
\usepackage{textcomp}
\usepackage{stfloats}
\usepackage{url}
\usepackage{verbatim}
\usepackage{graphicx}
\usepackage{orcidlink}
\usepackage{nomencl}
\usepackage{subcaption} % Keeping subcaption for subfigures, removed "caption" package as it's incompatible.
\usepackage[utf8]{inputenc}
\usepackage{pdfpages}
\usepackage{xcolor}
\usepackage{float}
\usepackage[letterpaper, left=0.673in, right=0.673in, top=0.767in, bottom=0.607in]{geometry}

\renewcommand\nomgroup[1]{%
  \item[\bfseries
  \ifstrequal{#1}{A}{Acronyms}{%
  \ifstrequal{#1}{M}{Mathematical Symbols}{}}%
]}

\def\BibTeX{{\rm B\kern-.05em{\sc i\kern-.025em b}\kern-.08em
    T\kern-.1667em\lower.7ex\hbox{E}\kern-.125emX}}

\begin{document}

\title{ViT Enhanced Privacy-Preserving Secure Medical Data Sharing and Classification}
\author{\IEEEauthorblockN{Al Amin\IEEEauthorrefmark{1},
        Kamrul Hasan\IEEEauthorrefmark{1},
        Sharif Ullah\IEEEauthorrefmark{2},
        M. Shamim Hossain\IEEEauthorrefmark{3}}\\
    \IEEEauthorblockA{\IEEEauthorrefmark{1}Department of Electrical \& Computer Engineering, College of Engineering, Tennessee State University, TN, USA}
    \IEEEauthorblockA{\IEEEauthorrefmark{2}Department of Computer Science \& Engineering, College of Science and Engineering, University of Central Arkansas, Conway, AR, USA}
    \IEEEauthorblockA{\IEEEauthorrefmark{3}Department of Software Engineering, College of Computer and Information Sciences, King Saud University, Riyadh, Saudi Arabia}\\
    Email: $\lbrace$\textit{aamin2, mhasan1}$\rbrace$@tnstate.edu, mullah@uca.edu, mshossain@ksu.edu.sa    
}

\maketitle

\begin{abstract}

Privacy-preserving and secure data sharing are critical for medical image analysis while maintaining accuracy and minimizing computational overhead are also crucial. Applying existing deep neural networks (DNNs) to encrypted medical data is not always easy and often compromises performance and security. To address these limitations, this research introduces a secure framework consisting of a \emph{learnable encryption method based on block-pixel operation} to encrypt the data and subsequently integrate it with the \emph{Vision Transformer (ViT)}. The proposed framework ensures data privacy and security by creating unique scrambling patterns per key, providing robust performance against leading bit attacks and minimum difference attacks.

% Privacy-preserving and secure data sharing is critical for medical image analysis, where maintaining accuracy and minimal computational overhead

\end{abstract}


\section{Introduction}
Advancements in ML and DL models, supported by cloud platforms like Google Cloud and Microsoft Azure, offer high computational power but also introduce privacy risks in shared environments \cite{DBLP:journals/corr/abs-2001-07761, 10273222,10296694}. In response, the research proposed ViT enhances the learnable encryption method and obfuscates image details while preserving classification-relevant features, ensuring HIPAA compliance. When tested on MRI and histopathological datasets, the method showed strong defense against attacks such as leading bit-attack, and minimum difference attacks, keeping medical data safe in the cloud.
\section{Proposed Approach}
\begin{figure}[h]
    \centering
    \includegraphics[height=4cm, width=\linewidth]{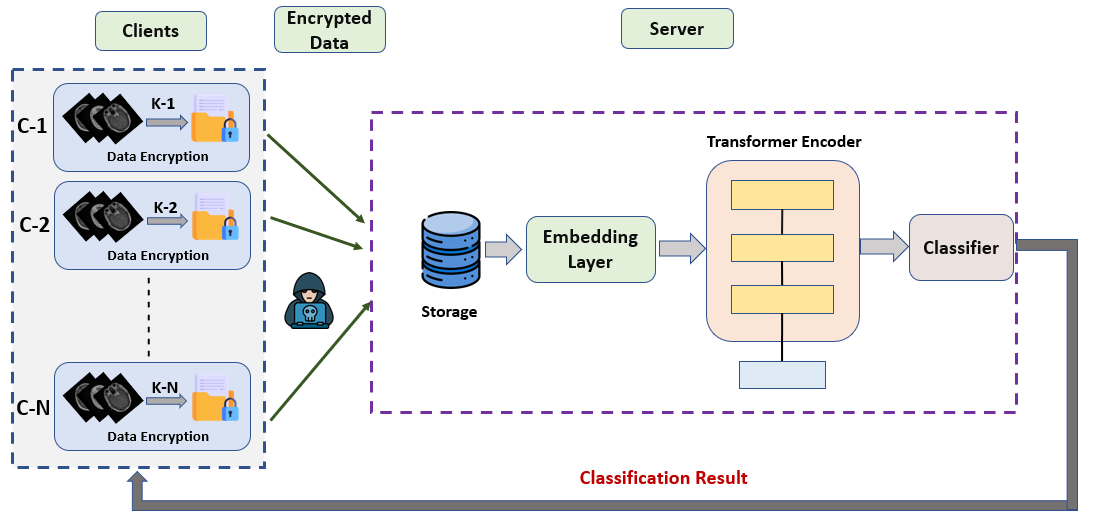}
    \caption{Overview of ViT integrated privacy-preserving secure medical data sharing and classification framework.}
    \label{fig:Pipeline}
\end{figure}

This research proposes ViT enhance privacy-preserving multi-layer learning approach for secure and robust medical image classification. Figure\ref{fig:Pipeline} shows the overview of the approach. Multiple clients (C-1, C-2, ..., C-N) encrypt their local images using distinct keys (K-1, K-2, ..., K-N), applying a learnable encryption technique based on Block-Pixel Operation, including negative-positive transformation and color channel shuffling, to obfuscate image details while preserving essential features. These encrypted datasets are transmitted to a central server and processed through an Embedding Layer that converts the scrambled patches into a high-dimensional space for the Transformer. The Transformer Encoder then captures complex dependencies across the encrypted patches, enabling robust feature extraction, and a classifier produces accurate medical image classifications. The proposed framework bridges various data owners' diverse feature spaces and builds a more efficient global model. It also considers potential adversarial attempts to reconstruct the encrypted data to the original data, ensuring the security of sensitive medical data in cloud-based environments.

\subsection{Learnable Encryption based on Block-Pixel Operation}
The Learnable Encryption technique secures medical images using block-wise scrambling, pixel and channel shuffling, and negative-positive transformations to obfuscate visual details effectively. The following section provides a mathematical representation of the components and operations of the proposed encryption system. 

%Despite these transformations, the method preserves essential features for accurate classification, ensuring robustness and privacy in medical image analysis.}

\begin{itemize}

    \item \textbf{Block-Wise Scrambling:}
    Each patch \( P_i \) is subjected to a scrambling operation, where the pixels within the patch are reordered based on a key \( K \). The scrambling operation is defined as:
    \begin{equation}
    S(P_i, K) = \text{Permute}(P_i, K)
    \end{equation}
    where \( S(P_i, K) \) is the scrambled patch, and \( \text{Permute}(P_i, K) \) is a function that reorders the pixels according to the key \( K \).

    \item \textbf{Shuffling Block Positions:}
    After scrambling, the positions of the patches are shuffled based on the key \( K \). Let the original patch positions be indexed as \( \{1, 2, \dots, 64\} \). The shuffling process is expressed as:
    \begin{equation}
    \text{Shuffle}(P_i, K) \rightarrow P_{\sigma(i)}
    \end{equation}
    where \( \sigma \) is a permutation of the patch indices determined by the key \( K \).

    \item \textbf{Negative-Positive Transformation:}
    Each pixel value in the scrambled patches undergoes a conditional inversion, where the pixel value \( x \) is transformed as follows:
    \begin{equation}
    x' = 
    \begin{cases} 
    x & \text{if } r = 0 \\
    255 - x & \text{if } r = 1 
    \end{cases}
    \end{equation}
    where \( r \) is a randomly generated binary integer (0 or 1) for each pixel, controlled by the key \( K \).

    \item \textbf{Color Channel Shuffling:}
    The color channels of each pixel (R, G, B) are shuffled based on the key \( K \). For a pixel with channels \( C = [R, G, B] \), the shuffled channels \( C' \) are given by:
    \begin{equation}
    C' = [C_{\sigma(1)}, C_{\sigma(2)}, C_{\sigma(3)}]
    \end{equation}
    where \( \sigma \) is a permutation of the channel indices \( \{1, 2, 3\} \) determined by the key \( K \).

    \item \textbf{Reconstruction of the Encrypted Image:}
    The final encrypted image \( I_{\text{encrypted}} \) is reconstructed by assembling the scrambled and transformed patches:
    \begin{equation}
    \begin{aligned}
       I_{\text{encrypted}} = &\ \text{Reassemble}(\{S(P_1, K), S(P_2, K), \dots, \\
    &\ S(P_{64}, K)\})
    \end{aligned}
    \end{equation}
    where \( \text{Reassemble} \) refers to reassembling the full image from its encrypted patches.
\end{itemize}

\subsection{Dataset Description and Model Performance}

Two medical image datasets were used \cite{Lung,msoud_nickparvar_2021}, 5,712 MRI images for brain tumor classification (with 1,311 images for testing) and 2,980 histopathological images for lung and colon cancer classification (with 311 for testing). As shown in Table  \ref{tab:performance_comparison}, the ViT model achieves higher training and validation accuracies of 95\% and 94\% on encrypted MRI brain tumor datasets, compared to the DNN models, which exhibit much lower accuracies ranging from 38\% to 51\%.

\begin{table*}[ht]
\centering
\caption{{Performance Comparison of DNN Models and the Proposed Framework on Plain and Encrypted dataset}}
\label{tab:performance_comparison}
\begin{tabular}{|c|c|c|cc|cc|cc|}
\hline
\ {Plain   Image} &
  \ {\begin{tabular}[c]{@{}c@{}}Encryption \\ Method\end{tabular}} &
  \ {Model} &
  \multicolumn{2}{c|}{Dataset} &
  \multicolumn{2}{c|}{Accuracy} &
  \multicolumn{2}{c|}{Computational   Time} \\ \cline{4-9} 
 &
   &
   &
  \multicolumn{1}{c|}{Name} &
  Type &
  \multicolumn{1}{c|}{\begin{tabular}[c]{@{}c@{}}training\\ accuracy\\ \% \end{tabular}} &
  \begin{tabular}[c]{@{}c@{}}validation\\ accuracy\\ \%\end{tabular} &
  \multicolumn{1}{c|}{\begin{tabular}[c]{@{}c@{}}training\\ time (m)\end{tabular}} &
  \begin{tabular}[c]{@{}c@{}}validation\\  time (m)\end{tabular} \\ \cline{2-9} 
 &
  N/A &
  ResNet50 &
  \multicolumn{1}{c|}{\begin{tabular}[c]{@{}c@{}}Brain\\ Tumor\end{tabular}} &
  MRI &
  \multicolumn{1}{c|}{94} &
  93 &
  \multicolumn{1}{c|}{17.28} &
  4.32 \\ \cline{2-9} 
 &
  N/A &
  \begin{tabular}[c]{@{}c@{}}MobileNet \\ V2\end{tabular} &
  \multicolumn{1}{c|}{\begin{tabular}[c]{@{}c@{}}Brain\\ Tumor\end{tabular}} &
  MRI &
  \multicolumn{1}{c|}{98} &
  96 &
  \multicolumn{1}{c|}{31.62} &
  7.9 \\ \hline
\ {Encrypted Image} &
  \ {\begin{tabular}[c]{@{}c@{}}pixel   \\ shuffling\end{tabular}} &
  ResNet50 &
  \multicolumn{1}{c|}{\begin{tabular}[c]{@{}c@{}}Brain\\ Tumor\end{tabular}} &
  MRI &
  \multicolumn{1}{c|}{46} &
  38 &
  \multicolumn{1}{c|}{541.6} &
  108.32 \\ \cline{3-9} 
 &
   &
  \begin{tabular}[c]{@{}c@{}}MobileNet \\ V2\end{tabular} &
  \multicolumn{1}{c|}{\begin{tabular}[c]{@{}c@{}}Brain\\ Tumor\end{tabular}} &
  MRI &
  \multicolumn{1}{c|}{51} &
  51 &
  \multicolumn{1}{c|}{184.92} &
  46.23 \\ \cline{3-9} 
 &
   &
  \begin{tabular}[c]{@{}c@{}}Inception \\ V3\end{tabular} &
  \multicolumn{1}{c|}{\begin{tabular}[c]{@{}c@{}}Brain\\ Tumor\end{tabular}} &
  MRI &
  \multicolumn{1}{c|}{48} &
  46 &
  \multicolumn{1}{c|}{432.36} &
  86.46 \\ \cline{2-9} 
 &
  \ {\textbf{\begin{tabular}[c]{@{}c@{}}Proposed \\ Encryption\end{tabular}}} &
  \ {\textbf{\begin{tabular}[c]{@{}c@{}} \\ ViT (Proposed Work)\end{tabular}}} &
  \multicolumn{1}{c|}{\textbf{Brain Tumor}} &
  \textbf{MRI} &
  \multicolumn{1}{c|}{\textbf{95}} &
  \textbf{94} &
  \multicolumn{1}{c|}{\textbf{82.25}} &
  \textbf{19.78} \\ \cline{4-9} 
 &
   &
   &
  \multicolumn{1}{c|}{Lung and clone} &
  \begin{tabular}[c]{@{}c@{}}Histopathological\end{tabular} &
  \multicolumn{1}{c|}{95} &
  94 &
  \multicolumn{1}{c|}{98.11} &
  24.53 \\ \hline
\end{tabular}
\end{table*}

\subsection{Security Evaluation}

\begin{figure}[h]
    \centering
    \includegraphics[width=\linewidth]{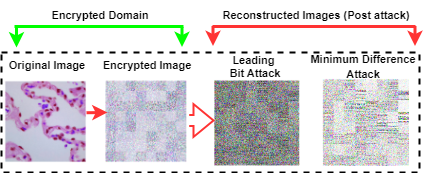}
    \caption{Visual Representation of Transformation Pipeline: Original, Encrypted, and Post-Attack Images.}
    \label{fig:attack_image}
\end{figure}

The proposed encryption method was rigorously evaluated against adversarial attacks, including leading bit attacks and minimum difference attacks. As illustrated in Figure \ref{fig:attack_image}, these attacks were unable to reconstruct the encrypted images, demonstrating the robustness of the model while maintaining high classification accuracy.

\section{Conclusion and Future Work}
This research introduces a secure Vision Transformer (ViT) model with learnable encryption for efficient medical image classification, outperforming traditional DNNs while protecting sensitive data. Future work will aim to scale the model, improve efficiency, and integrate federated learning for enhanced data privacy.

\bibliographystyle{ieeetr}
\bibliography{references.bib}
\end{document}